\documentclass[]{aa} 
\usepackage{times} 
\usepackage{psfig}
\def\bea{\begin{eqnarray}} 
\def\eea{\end{eqnarray}}

\def\xr{X--ray}
\def\xs{X--rays}
\def\sax{BeppoSAX}
\def\rosat{ROSAT}
\def\ginga{GINGA}
\def\exosat{EXOSAT}
\def\voyager{Voyager}
\def\vwh{VW~Hyi}
\begin{document}
\thesaurus{02.01.2, 08.09.2 \vwh, 08.14.2}
\title{The \xr\ spectra of VW Hydri during the outburst cycle}
\author{H.W.~Hartmann$^1$ \and  P.J.~Wheatley$^2$ \and J.~Heise$^1$ \and  
J.A.~Mattei$^3$ \and F.~Verbunt$^4$}
\authorrunning{H.W. Hartmann et al.}
\institute{$^1$SRON Laboratory for Space Research, Sorbonnelaan 2, NL-3584 CA 
Utrecht, the Netherlands \\
$^2$Department of Physics and Astronomy, University 
of Leicester, University Road, Leicester LE1 7RH \\
$^3$American Association of Variable Star Observers, 25 Birch Street, Cambridge 
USA, MA 02138-1205 \\
$^4$Astronomical Institute, P.O.Box 80000, NL-3508 TA 
Utrecht, the Netherlands}
\offprints{w.hartmann@sron.nl}
\date{Received ; accepted }
\maketitle
\begin{abstract}
We report six \sax\ \xr\ observations of \vwh\ during and after the outburst of 
Sep 23 1998. The outburst flux is lower than the quiescent flux in the entire 
observed energy band (0.1--10 keV), in agreement with earlier observations. The 
\xr\ spectra are fitted with two-temperature plasma and cooling flow spectral 
models. These fits show a clear spectral evolution in \xs\ for the first time in 
\vwh: the hard \xr\ turn-up after the outburst is reflected in the emission 
measure and the temperature. Moreover, during outburst the 1.5--10 keV flux 
decreases significantly. We argue that this is not consistent with the constant 
flux during a \rosat\ outburst observation made eight years earlier. We conclude 
from this observation that there are significant differences between outburst 
\xr\ lightcurves of \vwh.
\end{abstract}
\section{Introduction}
\label{intro}
\xs\ from dwarf novae arise very near the white dwarf, presumably in a 
boundary layer between the white dwarf and the accretion disk surrounding it.
Information on the properties of the \xr\ emitting gas as a function of the mass 
transfer rate through the accretion disk is provided by observations through the 
outburst cycle of dwarf novae. It may be hoped that such observations help to 
elucidate the nature of the \xr\ emission in cataclysmic variables, and by 
extension in accretion disks in general.
\par
\vwh\ is a dwarf nova that has been extensively studied during outbursts and 
in quiescence, at wavelengths from optical to hard \xs. It is a dwarf nova 
of the SU UMa type, i.e.\ in addition to ordinary dwarf nova outbursts it 
occasionally shows brighter and longer outbursts, which are called 
superoutbursts. Ordinary outbursts of \vwh\ occur every 20--30\,d and last 3--5 
days; superoutbursts occur roughly every 180\,d and last 10--14\,d (Bateson 
\cite{bateson77}).
\par
A multi-wavelength campaign combining data obtained with \exosat, \voyager, the 
International Ultraviolet Explorer, and by ground based optical observers 
covered three ordinary outbursts, one superoutburst, and the three quiescent 
intervals between these outbursts (Pringle et al. \cite{pringle87}, Van 
Amerongen et al. \cite{amerongen87}, Verbunt et al. \cite{verbunt87}, Polidan, 
Holberg \cite{polidan87}, van der Woerd \&\ Heise \cite{woerd87}). The \exosat\ 
data show that the flux in the 0.05--1.8\,keV range decreases during the 
quiescent interval; the flux evolution at lower energies and at higher energies 
(1--6\,keV) are compatible with this, but the count rates provided by \exosat\ 
are insufficient to show this independently. Folding the \exosat\ data of three 
outbursts showed that a very soft component appears early in the outbursts and 
decays faster than the optical flux (Wheatley et al. \cite{wheatley96}).
\par
The \rosat\ Position Sensitive Proportional Counter (PSPC) and Wide Field Camera 
(WFC) covered a dwarf nova outburst of \vwh\ during the \rosat\ All Sky Survey 
(Wheatley et al. \cite{wheatley96}). The PSPC data show that the flux in the 
0.1--2.5\,keV range is lower during outburst. The \rosat\ data showed no 
significant difference between outburst and quiescent \xr\ spectrum. The best 
spectral constraints are obtained for the quiescent \xr\ spectrum by combining 
\rosat\ WFC from the All Sky Survey with data from \rosat\ PSPC and \ginga\ 
pointings. A single temperature fit is not acceptable, the sum of two optically 
thin plasma spectra, at temperatures of 6\,keV and 0.7\,keV is somewhat better. 
The spectrum of a plasma which cools from 11\,keV and has emission measures at 
lower temperatures proportional to the cooling time, provides an acceptable fit 
of the spectrum in the 0.05--10\,keV energy range (Wheatley et al. 
\cite{wheatley96}).
\par
In this paper we report on a series of \sax\ observations of \vwh, which cover 
an ordinary outburst and a substantial part of the subsequent quiescent 
interval. The observations and data reduction are described in Sect.\,2, the 
results in Sect.\,3 and a discussion and comparison with earlier work is given 
in Sect.\,4.
\section{Observations and data reduction}
\label{obs}
\vwh\ is monitored at optical wavelengths by the American Association of 
Variable Star Observers (AAVSO). On Sep 23 1998 the optical magnitude of \vwh\ 
started to decrease. The outburst lasted for 5--6 days and reached a peak 
magnitude of 9.2. This outburst served as a trigger for a sequence of six 
observations by \sax\ between Sep 24 and Oct 18. As a result we have obtained 
one \xr\ observation during outburst and five observations during quiescence.
\par
\begin{figure*}[t] 
\psfig{file=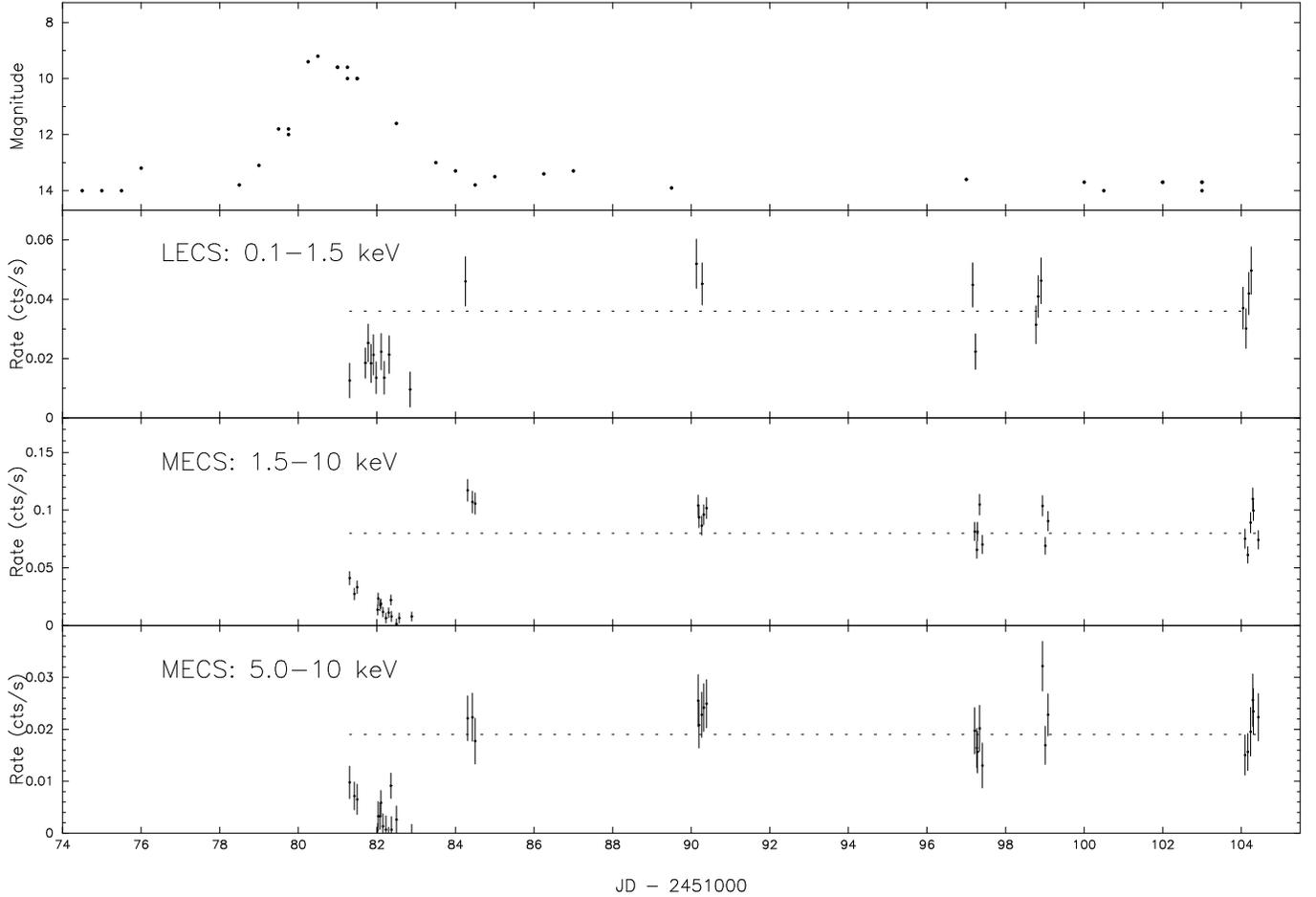,width=18cm} 
\caption{Optical and \xr\ lightcurves of \vwh. The \sax\ data have been 
accumulated in bins of 1024 and 1536 seconds for the LECS and the MECS 
respectively. The \sax\ MECS lightcurve is shown for the full energy range
1.5-10 keV, and also for the hard energy range only 5-10 keV. The six 
observation intervals can clearly be distinguished. The first interval coincides 
with the decline from the optical outburst. Indicated by the dotted lines are 
the average count rates of the combined observations 4--6} 
\label{lc}
\end{figure*}
Since \vwh\ appears as an on-axis source, the Low Energy Concentrator 
Spectrometer (LECS, Parmar et al. \cite{parmar97}) source counts are extracted 
from a circular region with a 35 pixel radius centered at the source. We use the 
Sep 1997 LECS response matrices centered at the mean raw pixel coordinates 
(130,124) for the channel-to-energy conversion and to fold the model spectra 
when fitted to the data. The combined Medium Energy Concentrator Spectrometer 
(MECS2 and MECS3, Boella et al. \cite{boella97}) source counts are extracted 
from a circular region with a 4\arcmin\ (30 pixel) radius. The September 1997 
MECS2 and MECS3 response matrices have been used. These matrices are added 
together. The background has been subtracted using an annular region with inner 
and outer radii of 35 and 49.5 pixels for the LECS and 30 and 42.5 pixels for 
the MECS, around the source region. 
\par
We ignore the data of the High Pressure Gas Scintillation Proportional Counter 
(HPGSPC, Manzo et al. \cite{manzo97}) and the Phoswitch Detection System (PDS, 
Frontera et al. \cite{frontera97}) since their background subtracted spectra 
have a very low signal to noise ratio.
\par
The LECS and MECS data products are obtained by running the \sax\ Data Analysis 
System pipeline (Fiore et al. \cite{fiore99}). We rebin the energy 
channels of all four instruments to ${1 \over 3}\times \mbox{FWHM}$ of the 
spectral resolution and require a minimum of 20 counts per energy bin to allow 
the use of the chi-squared statistic. The total LECS and MECS net exposure times 
are 82.5 ksec. and 181.4 ksec. respectively. The factor 2.2 between the LECS and 
MECS exposure times is due to non-operability of the LECS on the daytime side of 
the earth.
\section{Results}
\begin{table}[t] 
\caption[]{Observation dates, exposure times and background subtracted count 
rates for the \sax\ LECS (0.1--10 keV) and MECS (1.5--10 keV)}
\label{table1}
\begin{flushleft} 
\begin{tabular}{crrrrr} 
\hline\noalign{\smallskip}
\multicolumn{1}{c}{Obs.} & \multicolumn{1}{c}{Obs. date} & 
\multicolumn{2}{c}{LECS} & \multicolumn{2}{c}{MECS} \\
 & & \multicolumn{1}{c}{${\rm t_{exp}}$} & \multicolumn{1}{c}{cnt. rate} & 
\multicolumn{1}{c}{${\rm t_{exp}}$} & \multicolumn{1}{c}{cnt. rate} \\
 & & \multicolumn{1}{c}{ksec} & \multicolumn{1}{c}{cts s$^{-1}$} & 
\multicolumn{1}{c}{ksec} & \multicolumn{1}{c}{cts s$^{-1}$} \\
\noalign{\medskip}
\hline\noalign{\smallskip}
 1 & 24--26/09/1998 & 34.1 & 0.024(3)  & 76.6 & 0.016(2) \\
 2 & 27--28/09/1998 &  7.2 & 0.098(8)  & 20.2 & 0.109(8) \\
 3 &   3--4/10/1998 &  9.1 & 0.089(8)  & 17.4 & 0.094(6) \\
 4 & 10--11/10/1998 &  9.4 & 0.072(7)  & 20.0 & 0.075(5) \\
 5 &     12/10/1998 & 11.1 & 0.072(6)  & 21.6 & 0.085(5) \\
 6 & 17--18/10/1998 & 11.6 & 0.077(6)  & 25.6 & 0.078(5) \\
\noalign{\medskip}
\hline\noalign{\smallskip}
Total   &                & 82.5 & & 181.4 & \\
\noalign{\medskip}
\hline 
\end{tabular} 
\end{flushleft} 
\end{table}
\begin{figure}[t] 
\psfig{file=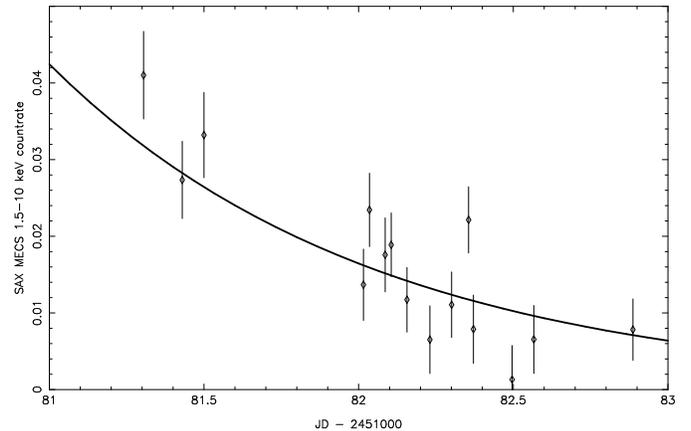,width=8.8cm} 
\caption{Close-up of the MECS observation 1 lightcurve. The solid curve shows 
the fitted exponential decay} 
\label{decay}
\end{figure}
In Fig. \ref{lc} we show the optical lightcurve, provided by the {\em American 
Association of Variable Star Observers} and the {\em Variable Star Network}, of 
\vwh\ at the time of our \xr\ observations. These optical observations show 
that our first \sax\ observation was obtained during an ordinary outburst that 
peaked on Sep 24, whereas observations 2--6 were obtained in quiescence. The 
last ordinary outbursts preceding our first \sax\ observation was observed by 
the AAVSO to peak on Sep 8; the first outburst observed after our last \sax\ 
observation was a superoutburst that started on Nov 5 and lasted until Nov 19.
\subsection{Lightcurve}
\label{rate_evolution}
In Fig. \ref{lc} we also show the count rates detected with the \sax\ LECS and 
MECS. For the latter instrument we show the count rates separately for the full 
energy range 1.5-10 keV, and for the hard energies only in the range 5-10 keV. 
In both LECS and MECS the count rate is lower during the outburst than in 
quiescence. In quiescence the count rate decreases significantly between our 
second and third (only in the MECS data), and between the third and fourth 
observations (both LECS and MECS data), but is constant after that (see Table 
\ref{table1}).
\par
The MECS count rate decreases during our first observation, when \vwh\ was in 
outburst, as is shown in more detail in Fig. \ref{decay}. This decrease can be 
described as exponential decline $N_{\rm ph}\propto e^{-t/\tau}$ with 
$\tau\simeq 1.1$\,d. The count rates in the LECS are compatible with the same 
decline, but the errors are too large for an independent confirmation. The 
count rates at lower energies, 0.1--1.5\,keV, are compatible with both a 
constant value and the exponential decay during our first observation.
\subsection{Spectral fits}
\begin{figure}[t] 
\psfig{file=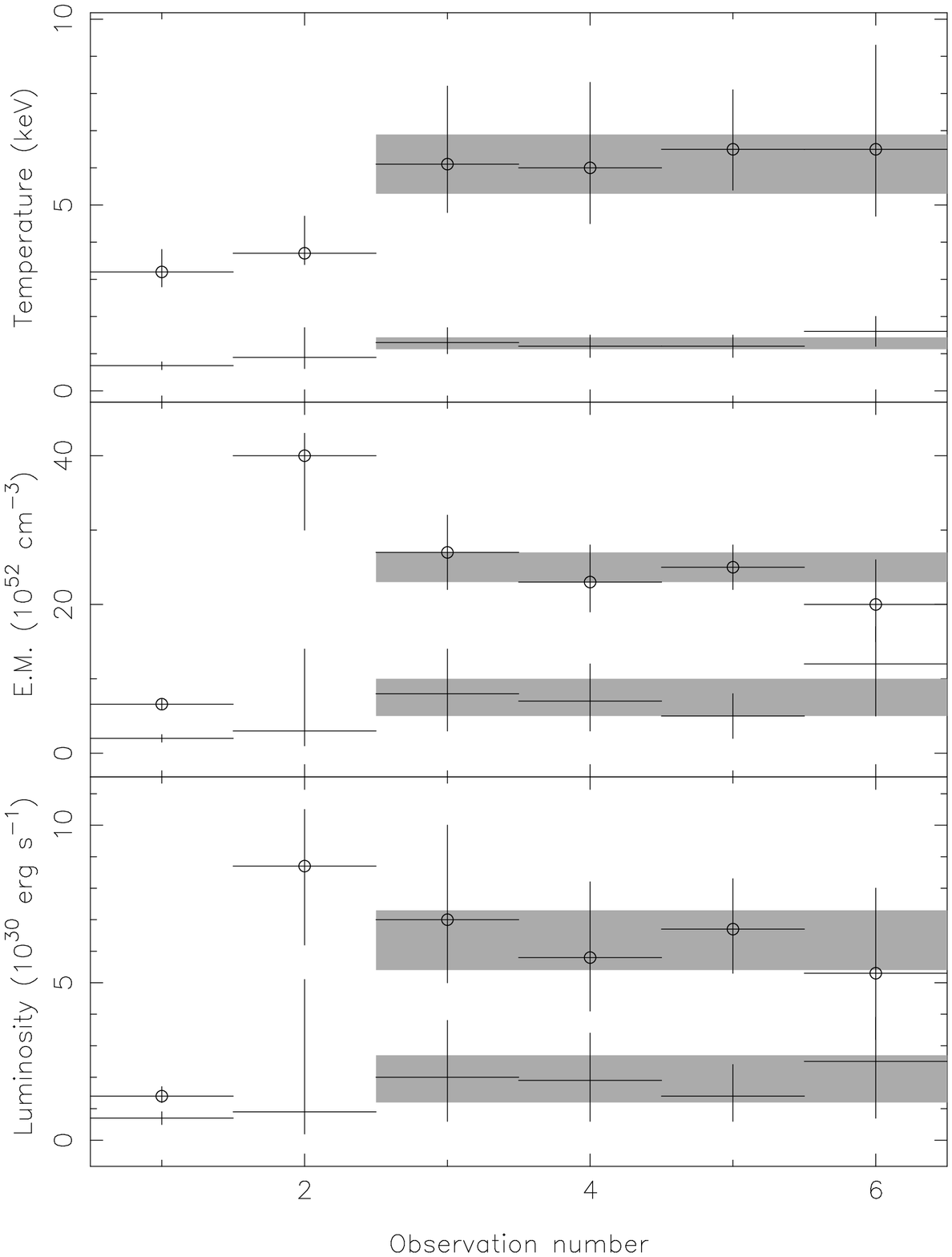,width=8.8cm} 
\caption{{\bf a--c} Fit parameters with 90\% confidence intervals for a 
two-temperature plasma model plotted against the observation number. The hot 
component is indicated by open circles. The parameter ranges for the spectrum of 
the combined observations 3--6 are indicated by the grey areas} 
\label{param}
\end{figure}
\begin{figure}[t] 
\psfig{file=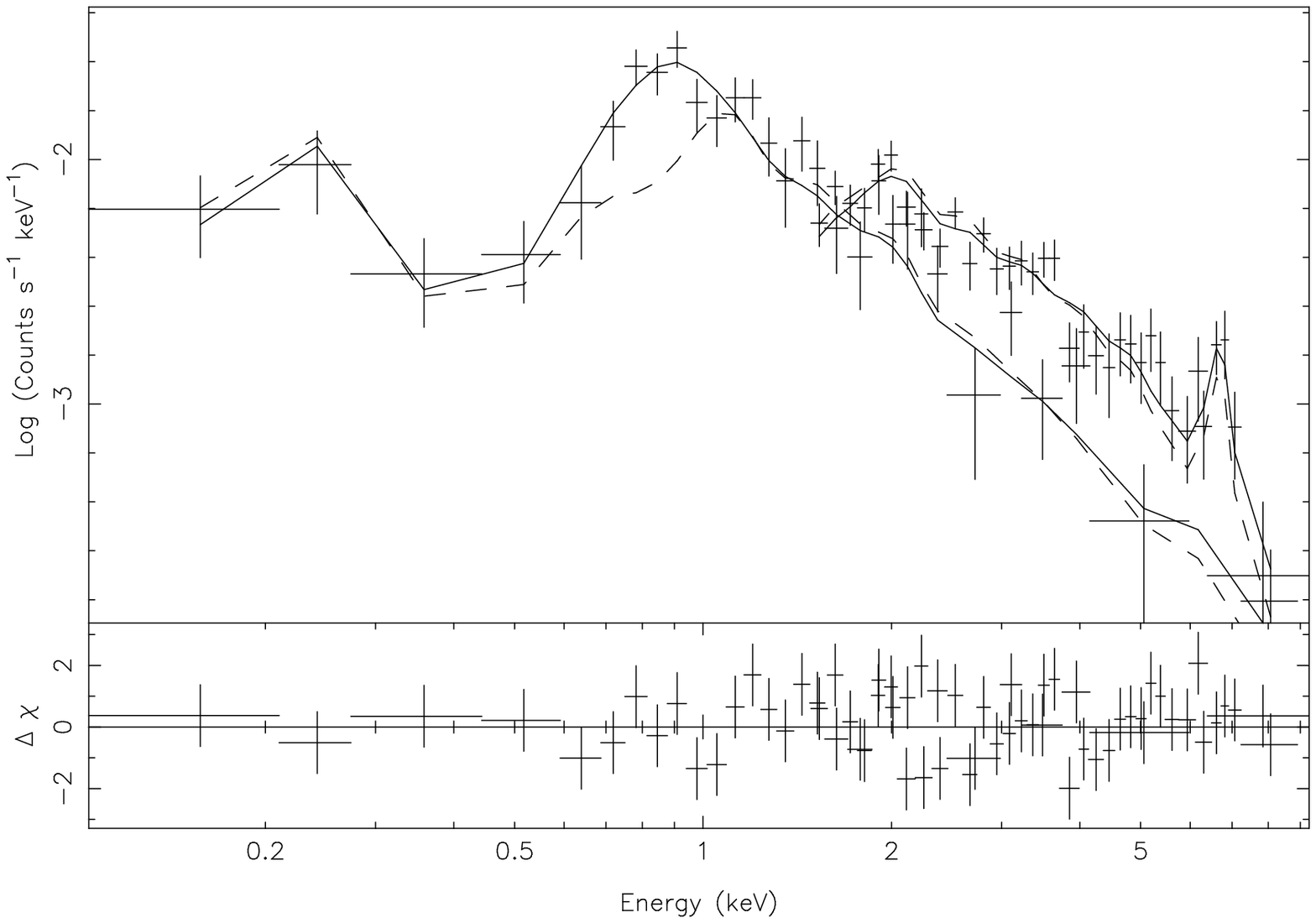,width=8.8cm}\par
\psfig{file=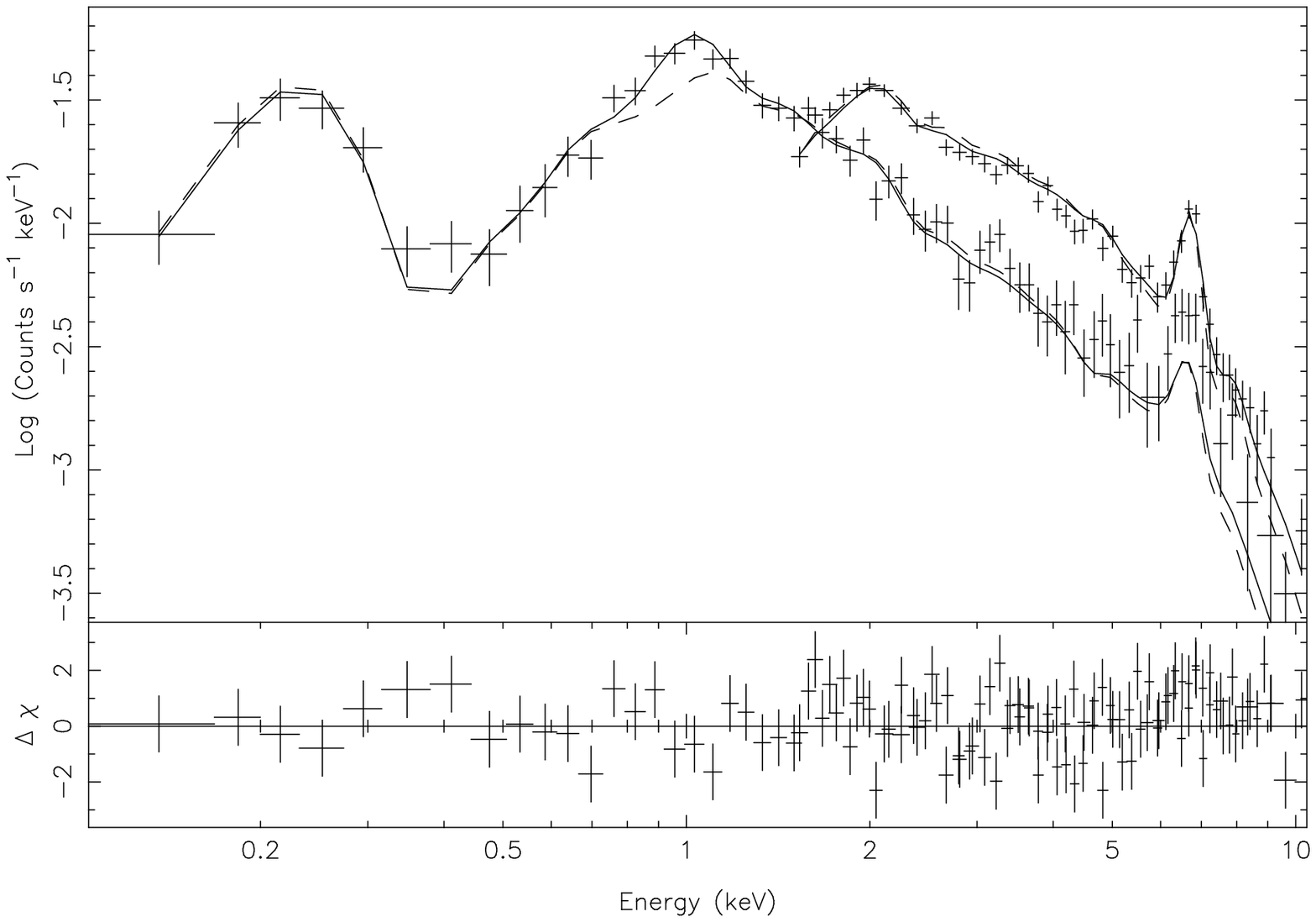,width=8.8cm}
\caption{Count rate spectra of the LECS/MECS observation 1 (top panel) and of 
combined observations 3-6  (lower panel). The best spectral fits for a 
single-temperature spectrum and for a two-temperature spectrum are shown as 
dashed lines and solid lines, respectively. The excess due to the Fe-L emission 
line complex is made visible in the one-component fit. This excess is filled up 
by adding a second, cooler, plasma component. The residuals of the two-component 
fits are indicated as well}
\label{2comp_err}
\end{figure}
\begin{figure}[t] 
\psfig{file=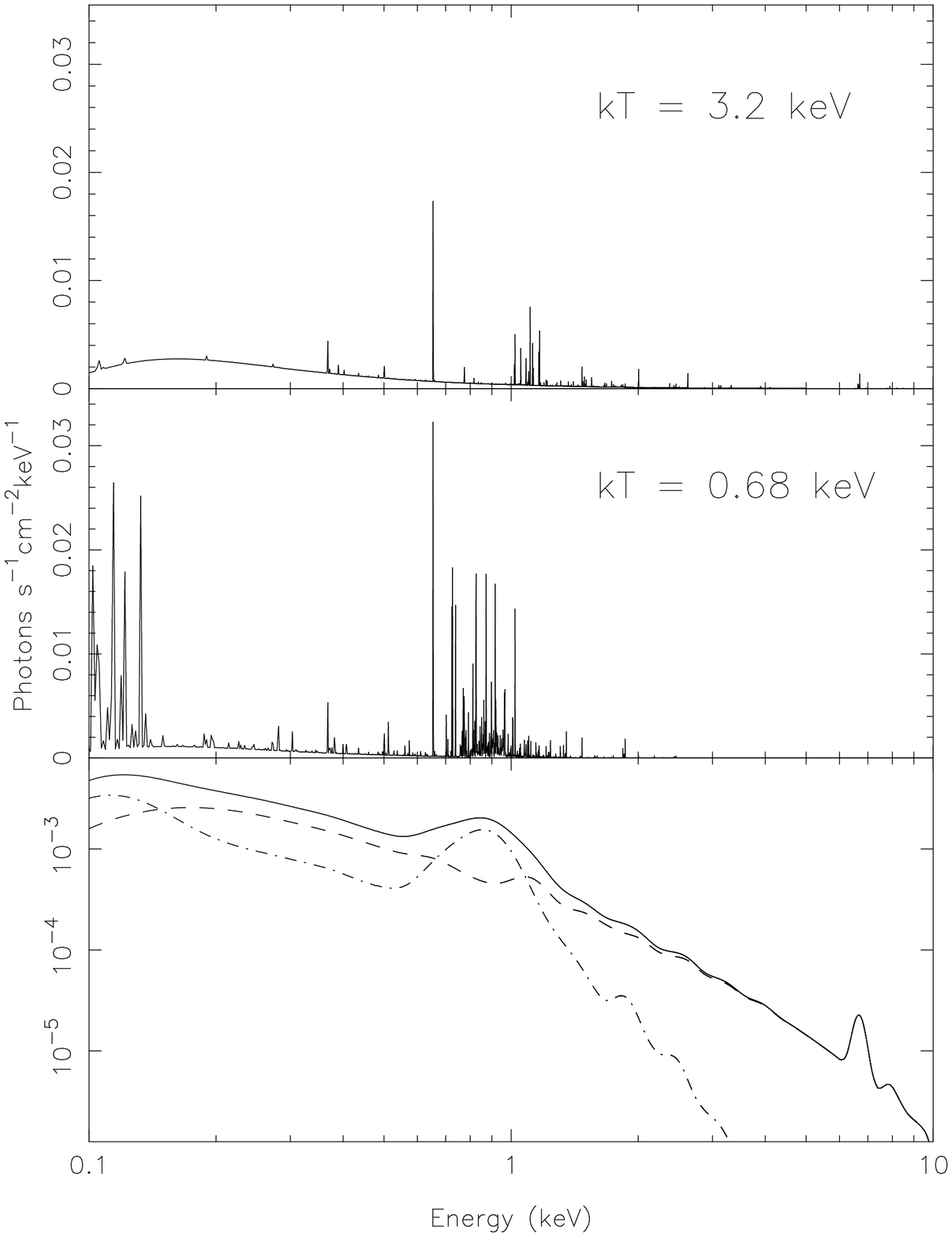,width=8.8cm} 
\caption{In the top and middle panel are plotted the hot and cool 
plasma components used in the fit of observation 1, on a linear scale. In the 
bottom panel are plotted, on a logarithmic scale, the high (dashed line) and low 
(dashed-dotted line) temperature photon spectra and their sum (solid line), 
folded with a Gaussian representing the \sax\ spectral response function. This 
demonstrates that the Fe-M line emission near 0.1 keV and the Fe-L line emission 
near 1 keV of the cool component contributes significantly to the total photon 
spectrum} \label{smooth}
\end{figure}

\begin{table*}[t]
\caption{Fit results for a two-temperature plasma model and for a cooling flow 
model. The errors indicated are the 90\% confidence intervals. The emission 
measures and luminosities have been calculated assuming a distance to \vwh\ of 
65 pc (see Warner \cite{warner87})}
\label{fitres2}
\begin{flushleft}
\begin{tabular}{cllllllll}
\noalign{\smallskip}
\hline
\noalign{\smallskip}
\multicolumn{9}{c}{\bf Two-temperature} \\
\noalign{\medskip}
Obs. & T$_1$ & T$_2$ & E.M.$_1$ & E.M.$_2$ & L$_1$ & L$_2$ & 
${\rm n_H}$ & $\chi^2$ (d.o.f.) \\
 & keV & keV & $10^{52}\mbox{ cm}^{-3}$ & $10^{52}\mbox{ cm}^{-3}$ & 
$10^{30} \mbox{erg s}^{-1}$ & $10^{30} \mbox{erg s}^{-1}$ & 
$10^{19}\mbox{cm}^{-2}$ & \\
\noalign{\medskip}
\hline
\noalign{\smallskip}
 1 & $0.68_{-0.11}^{+0.10}$ & $3.2_{-0.4}^{+0.6}$ & $2.0_{-0.5}^{+0.5}$ & 
$6.6_{-0.7}^{+0.7}$ & $0.7_{-0.2}^{+0.2}$ & $1.4_{-0.2}^{+0.3}$ & $4^*$  & 65 
(61) \\
 2 & $ 0.9_{-0.3}^{+0.8}$   & $3.7_{-0.3}^{+1.0}$ & $3_{-2}^{+11}$ & 
$40_{-10}^{+3}$      & $1.0_{-0.7}^{+4.2}$ & $9_{-3}^{+2}$ & $4^*$   & 92 (70) 
\\
 3 & $ 1.3_{-0.3}^{+0.4}$   & $6.1_{-1.3}^{+2.1}$ & $8_{-5}^{+6}$       & 
$27_{-5}^{+5}$      & $2.0_{-1.4}^{+1.8}$ & $7_{-2}^{+3}$       & 
$4^*$    & 103 (75) \\
 4 & $ 1.2_{-0.3}^{+0.3}$   & $6.0_{-1.5}^{+2.3}$ & $7_{-4}^{+5}$       & 
$23_{-4}^{+5}$      & $1.9_{-1.3}^{+1.5}$ & $6_{-2}^{+2}$ & $4^*$    & 75 (66) 
\\
 5 & $ 1.2_{-0.3}^{+0.3}$   & $6.5_{-1.1}^{+1.6}$ & $5_{-3}^{+3}$       & 
$25_{-3}^{+3}$      & $1.4_{-0.8}^{+1.0}$ & $6.7_{-1.4}^{+1.6}$ & $4^*$   & 84 
(74) \\
 6 & $ 1.6_{-0.4}^{+0.4}$   & $6.5_{-1.8}^{+2.8}$ & $12_{-7}^{+5}$      & 
$20_{-5}^{+6}$      & $2.5_{-1.8}^{+1.4}$ & $5_{-2}^{+3}$       & $4^*$ 
    & 78 (77) \\
\noalign{\medskip}
\hline
\noalign{\smallskip}
 3--6   & $1.28_{-0.16}^{+0.16}$ & $6.0_{-0.7}^{+0.9}$ & $7_{-2}^{+3}$ & 
$25_{-2}^{+2}$ & $1.9_{-0.7}^{+0.8}$ & $6.3_{-0.9}^{+1.0}$ & $4_{-2}^{+3}$ & 
149 (113) \\
\noalign{\medskip}
\hline
\noalign{\smallskip}
\multicolumn{9}{c}{\bf Cooling flow} \\
\noalign{\medskip}
Obs. & ${\rm T}_{\rm low}$ & ${\rm T}_{\rm high}$ & 
\multicolumn{2}{l}{Normalization} & \multicolumn{2}{l}{L} & ${\rm n_H}$ 
 & $\chi^2$ (d.o.f.) \\
 & keV & keV & \multicolumn{2}{l}{$10^{-12}\mbox{ M}_\odot\mbox{yr}^{-1}$} 
 & \multicolumn{2}{l}{$10^{30} \mbox{erg s}^{-1}$} & $10^{19}\mbox{cm}^{-2}$ & 
\\
\noalign{\medskip}
\hline
\noalign{\smallskip}
 1 & $<0.4$ & $4.5^{+0.4}_{-0.6}$ & \multicolumn{2}{l}{$1.8^{+0.2}_{-0.2}$} & 
\multicolumn{2}{l}{$1.7^{+0.2}_{-0.2}$} & $4^*$ & 76 (62)   \\
 2 & $1.0^{+0.5}_{-0.3}$ & $6.8^{+1.1}_{-0.5}$ & 
\multicolumn{2}{l}{$6.0^{+0.2}_{-0.5}$} & 
\multicolumn{2}{l}{$8.5^{+1.7}_{-0.8}$} & $4^*$ & 92 (71)   \\
 3--6 & $0.66^{+0.18}_{-0.08}$ & $9.9^{+0.8}_{-0.8}$ & 
\multicolumn{2}{l}{$3.1^{+0.3}_{-0.1}$} & 
\multicolumn{2}{l}{$7.2^{+0.6}_{-1.2}$} & $4^*$   & 153 (115) \\
\noalign{\medskip}
\hline
\end{tabular}
\end{flushleft}
\noindent{$^*$\,Fixed parameter c.f. the value obtained from the combined 
observations 3--6}
\end{table*}
We have made spectral fits to the combined MECS and LECS data for each of the 
six separate \sax\ observations and computed the luminosities assuming a 
distance of 65 pc to \vwh\ (see Warner \cite{warner87}). As expected on the 
basis of earlier work, described in the introduction, we find that the observed 
spectra cannot be fitted with a single-temperature plasma. The combination of 
spectra of optically thin plasmas at two different temperatures does provide 
acceptable fits. The parameters of these fits are listed in Table \ref{fitres2}, 
and their variation between the separate observations is illustrated in Fig. 
\ref{param}. The need for a two-temperature fit is illustrated in Figs. 
\ref{2comp_err} and \ref{smooth} for the outburst spectrum of observation 1 and 
for the quiescent spectrum of the combined observations 3--6: the low 
temperature component is required to explain the excess flux near 1 keV. The 
Fe-K emission line near $6.70\pm 0.05\mbox{ keV}$ is clearly present in our 
data, and is due to hydrogen or helium like iron from the hot component of the 
plasma. The LECS data in observations 3--6 are poorly fitted above $\sim 5$ keV 
which is probably due to calibration uncertainties of the instrument (Fiore et 
al. \cite{fiore99}). We fix $n_{\rm H}$ at $4\times 10^{19}\mbox{ cm}^{-2}$, the 
best-fit value of the combined observation 3--6. (Fixing $n_{\rm H}$ at $6\times 
10^{17}\mbox{ cm}^{-2}$, which was found by Polidan et al. (\cite{polidan90}), 
does not change the fit parameters, except for the chi-squared values of 
observations 2, 3 and 3--6 which become slightly worse; 98, 111 and 158 
respectively.)
\par
The temperature of both the cool and the hot component of the two-temperature 
plasma is higher during quiescence than during the outburst, increasing from 
respectively 0.7\,keV and 3.2\,keV in outburst to 1.3\,keV and 6\,keV in 
quiescence. The temperatures immediately after outburst -- in our second  
observation -- are intermediate between those of outburst and quiescence. The 
emission measure (i.e.\ the integral of the square of the electron density over 
the emission volume, $\int n_{\rm e}^2dV$) of both the cool and the hot 
component of the two-temperature plasma is also higher in quiescence; 
immediately after outburst the emission measure of the hot component is higher 
than during the later phases of quiescence. The temperatures and emission 
measures of the two-temperature plasma are constant, within the errors, in the 
later phases of quiescence of our observations 3--6. For that reason, we have 
also fitted the combined data of these four observations to obtain better 
constraints on the fit parameters (see Table \ref{fitres2}). Note that the 
decrease of the count rate between observations 3 and 4, mentioned in Sect. 
\ref{rate_evolution}, is significant even though it is not reflected in the 
emission measures and luminosities of the two components separately. This is due 
to the combined spectral fitting of the LECS and the MECS, since the decrease in 
count rate is less significant for the LECS. Moreover, the errors on the count 
rates are much smaller than those on the emission measures ($\la 10\%\mbox{ and 
}\ga 20\%$ respectively).
\par
\begin{table}[t] 
\caption[]{The spectral parameters for the first 31 ksec and next 46 ksec of 
the outburst spectrum. From these values the MECS and \rosat\ PSPC count rates 
are predicted}
\label{vwh_splitobs1}
\begin{flushleft} 
\begin{tabular}{cllll} 
\hline\noalign{\smallskip}
 Obs. & T$_2$ & E.M.$_2$ & MECS & PSPC \\
 & keV & $10^{52}\mbox{ cm}^{-3}$ & cts s$^{-1}$ & cts s$^{-1}$ \\
\noalign{\medskip}
\hline\noalign{\smallskip}
1a & $3.6_{-0.7}^{+1.3}$ & $8.4_{-1.3}^{+1.3}$ & $0.022_{-0.003}^{+0.003}$ & 
$0.35_{-0.03}^{+0.03}$ \\
1b & $3.0_{-0.6}^{+1.3}$ & $5.4_{-0.8}^{+0.8}$ & $0.013_{-0.002}^{+0.002}$ & 
$0.23_{-0.02}^{+0.02}$ \\
\noalign{\medskip}
\hline
\end{tabular} 
\end{flushleft} 
\end{table}
We fit the first 31 ksec and the next 46 ksec of the outburst spectrum (1a and 
1b) separately. Both fits are good with $\chi^2<1$. From the fit results we 
compute the MECS and \rosat\ PSPC count rates. The results are shown in Table 
\ref{vwh_splitobs1}. We have only indicated the temperature and emission measure 
of the hot component since the cool component is responsible for the iron line 
emission outside the MECS bandwidth and does not have a large impact upon the 
continuum emission. Note from Table \ref{vwh_splitobs1} that the decay in count 
rate is entirely due to the decrease of the emission measure.
\par
To compare our observations with the results obtained by Wheatley et al. 
(\cite{wheatley96}) we consider next the cooling flow model (cf. Mushotzky, 
Szymkowiak \cite{mushotzky88}) for our observations 1, 2 and 3--6. In this model 
the emission measure for each temperature is restricted by the demand that it is 
proportional to the cooling time of the plasma. The results of the fits are 
shown in Table \ref{fitres2}. Note that these results are not better than the 
two-temperature model fits. Due to the poor statistics of the LECS outburst 
observation we cannot constrain the lower temperature limit. The MECS is not 
sensitive to this temperature regime at all. A contour plot of the upper and 
lower temperature limits for the combined quiescent observations 3--6 is shown 
in Fig. \ref{contour36}. The boundaries of the low temperature in Fig. 
\ref{contour36} are entirely determined by the Fe-L and Fe-M line emission; for 
a low temperature of $\la0.35$ keV the contributions to the line flux integrated 
over all higher temperatures exceeds the observed line flux. For a low 
temperature of $\ga1.2$ keV there is not sufficient line flux left in the model. 
The boundaries of the high temperature are determined by the continuum slope; 
for a high temperature of $\la8.5$ and $\ga11.5$ keV the model spectrum is too 
soft and too hard respectively to fit the data.
\section{Comparison with previous \xr\ observations}
\label{vwh_discussion}
\subsection{Time variability}
\label{vwh_timevar}
\begin{figure}[t]
\psfig{file=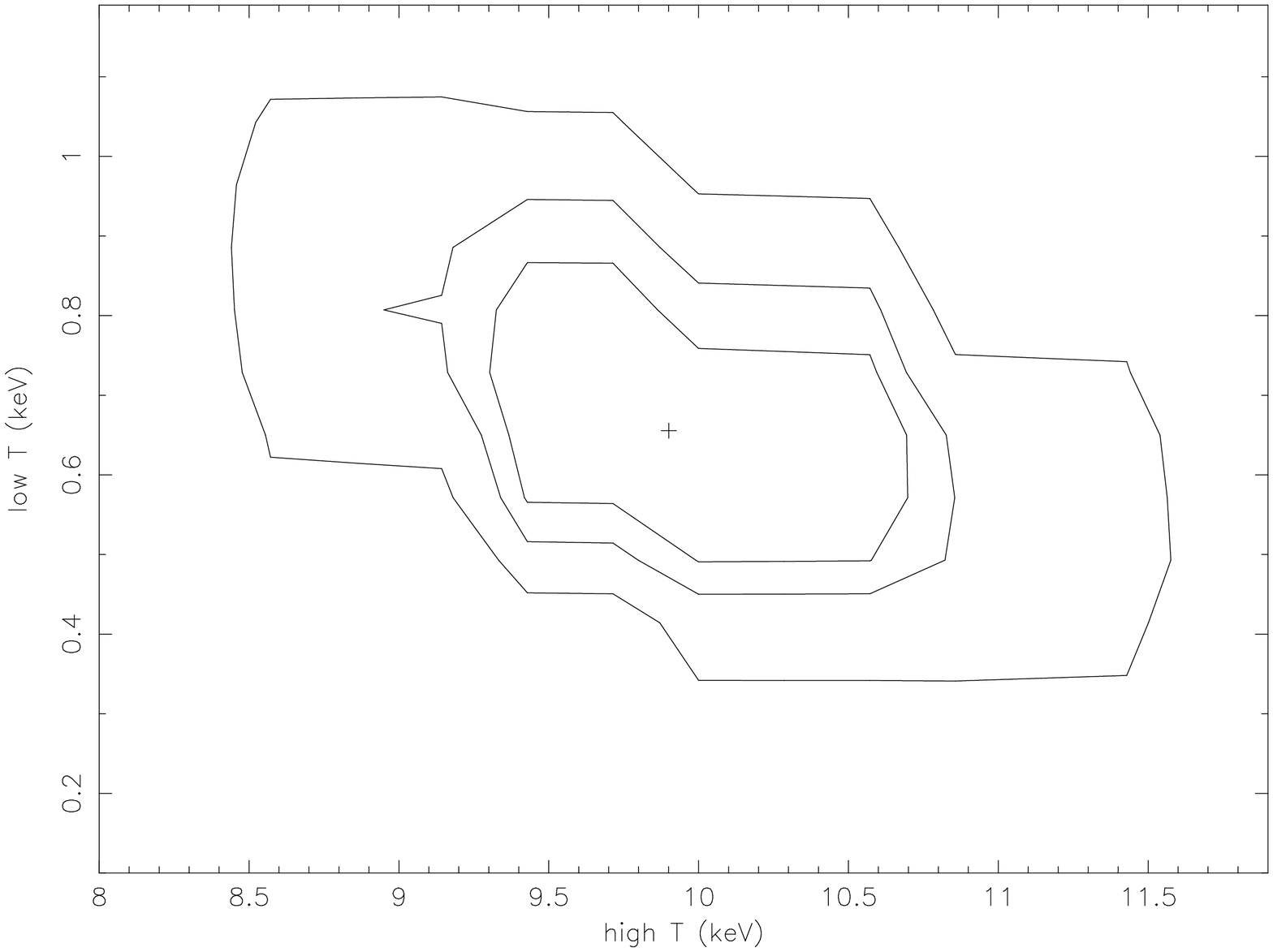,width=8.8cm}
\caption{Confidence contours plotted as a function of the upper and lower 
temperature for a cooling plasma during quiescence. See text. The contours 
represent the 68, 90 and 99\% confidence levels of the fit to the observations 
3--6}
\label{contour36}
\end{figure}
We predict the \rosat\ count rates of \vwh\ during outburst and quiescence with 
the observed \sax\ flux from the two-temperature fit (see Table \ref{fitres2}). 
Here we do apply $n_{\rm H}=6\times 10^{17}\mbox{ cm}^{-2}$ (Polidan et al. 
\cite{polidan90}) since \rosat\ is probably more sensitive to $n_{\rm H}$ than 
\sax. The predicted count rates during outburst and quiescence are 0.31 and 
$0.87\mbox{ cts s}^{-1}$. The \rosat\ observed count rates are 0.4 and 
$1.26\mbox{ cts s}^{-1}$ respectively (Belloni et al. \cite{belloni91}; Wheatley 
et al. \cite{wheatley96}). Both predictions appear to be different from the 
observations by a factor $\sim0.75$.
\par
From Fig. \ref{decay}, we observe a decrease in MECS count rate by a factor of 
$\ga 4$ during outburst. This is inconsistent with the constant 0.4 $\mbox{ 
cts s}^{-1}$ observed by the \rosat\ PSPC during outburst (Wheatley et al. 
\cite{wheatley96}). Using the LECS data during the outburst in a bandwidth 
(0.1--1.5 keV) comparable to the \rosat\ PSPC we cannot discriminate 
observationally between a constant flux and the exponential decay observed by 
the MECS. However, our spectral fits to the data require that the 0.1-2.5 keV 
flux decreases in tandem with the hard flux. Thus the difference between the 
\rosat\ PSPC and the \sax\ MECS lightcurves during outburst may either be due to 
variations between individual outbursts or to the different spectral bandwidths 
of the observing instruments. The predicted decay of the count rate 
significantly exceeds the range allowed by the ROSAT observations of the Nov 
1990 outburst.
\par
We interpret the time variability of the count rate shown in Figs. \ref{decay} 
and \ref{param}, as a change mainly in the amount of gas in the inner disk that 
emits keV photons. At the end of the outburst, while the inner disk is still 
predominantly optically thick, the mass accretion rate onto the white dwarf is 
decreasing. As a result, the amount of hot optically thin gas drops gradually. 
This is observed in Fig. \ref{decay}. The transition to a predominantly 
optically thin inner disk  occurs just before observation 2. As a result the 
amount of optically thin emitting material in the disk increases strongly. This 
is shown by the increase of the emission measure of the hot component in Fig. 
\ref{param}, observation 2, which even peaks above the quiescent value. The 
settling of the accretion rate towards quiescence is shown in Fig. \ref{param}, 
observations 3--6 for both the temperature and the emission measure. In contrast 
to the emission measure, the temperature of the hot component increases only 
gradually throughout observations 1--6 as it reflects the slowly decreasing 
accretion rate rather than the amount of optically thin emitting material in the 
disk.
\subsection{Spectral variability}
\label{vwh_specvar}
Both a two-temperature plasma model and a cooling flow model fit the spectrum 
of our \sax\ observations of \vwh\ better than a one-temperature model. The 
contribution of the cool component lies mainly in the presence of strong 
Fe-L line emission around 1 keV. The hot component contributes the continuum 
and the Fe-K line emission at $\sim 6.7$ keV. Adding a soft atmospheric 
component in the form of a $\la 10$ eV  blackbody model does not improve our 
fits. This blackbody component, reported by Van~der~Woerd et al. 
(\cite{woerd86}) and Van~Teeseling et al. (\cite{teeseling93}), is too soft to 
be 
detected by \sax\ LECS.
\par
Based upon the $\chi^2$-values, the \sax\ observation of \vwh\ does not 
discriminate between a continuous temperature distribution (the cooling flow 
model) and a discrete temperature distribution (the two-component model) of the 
\xr\ emitting region. Wheatley et al. (\cite{wheatley96}) derive a lower and 
upper 
temperature of $\la 0.53\mbox{ and }11^{+3}_{-2}$ keV respectively for a 
cooling flow fit to the combined \rosat\ PSPC and \ginga\ LAC data during 
quiescence. These temperatures are consistent with our cooling flow fits to 
\sax\ data during quiescence; there is a small overlap between the 2 and 
3$\sigma$ contours shown in Fig. 6 by Wheatley et al. and the contours of our 
Fig. \ref{contour36}.
\section{Conclusions}
\label{vwh_conclusions}
\sax\ does not discriminate between a continuous (cooling flow) and a discrete 
temperature distribution. Our observation of a decreasing count rate, followed 
by a constant count rate during quiescence is in contradiction with the disk 
instability models. These models predict a slightly increasing mass transfer 
onto the white dwarf which must show up as an {\em increase} in the \xr\ flux. 
{\em Ad hoc} modifications to disk instability models, such as interaction of 
the inner disk with a magnetic field of the white dwarf (Livio, Pringle 
\cite{livio92}), evaporation of the inner disk (Meyer, Meyer-Hofmeister 
\cite{meyer94}), or irradiation of the inner disk by the white dwarf (King 
\cite{king97}), possibly are compatible with the decrease of ultraviolet flux 
(e.g. Van Amerongen et al. \cite{amerongen90}) and \xr\ flux during quiescence.
\par
If we assume a continuous temperature distribution the upper temperature limit 
of our quiescence spectrum is consistent with the observations by Wheatley et 
al. (\cite{wheatley96}). The cooling flow model requires an accretion rate of 
$3\times 10^{-12}\mbox{ M}_\odot\mbox{ yr}^{-1}$ to explain the \xr\ luminosity 
late in quiescence. A similar result is obtained when we convert the luminosity 
derived from the two-temperature model to an accretion rate. Any outburst model 
must accommodate this accretion rate.
\par
\sax\ MECS observes a significant decrease in the count rate during outburst. 
Our simulations show a similar decrease for the \rosat\ PSPC which would have 
been significantly detected. The fact that the \rosat\ count rate during 
outburst was constant (Wheatley et al. \cite{wheatley96}) and the results from 
our cooling flow model fits suggest that the outburst of Sep 24 1998 behaved 
differently from the outburst of Nov 3 1990.
\begin{acknowledgements}
This work has been supported by funds of the Netherlands Organization for 
Scientific Research (NWO). 
\end{acknowledgements} 
\end{document}